\RequirePackage{fix-cm}
\documentclass[smallextended]{svjour3}       % onecolumn (second format)
\smartqed  % flush right qed marks, e.g. at end of proof
\usepackage{graphicx}
\usepackage{mathptmx}      % use Times fonts if available on your TeX system
\usepackage{latexsym}
\usepackage{color}
\usepackage{eurosym}
\usepackage{braket}
\usepackage{amsmath}
\usepackage{amssymb}
\usepackage[hyphens]{url}
\PassOptionsToPackage{hyphens}{url}\usepackage{hyperref}
\journalname{Ethics and Information Technology}
\begin{document}

\title{On the impact of quantum computing technology on future developments in high-performance scientific computing}

\titlerunning{Impact of quantum computing on future scientific computing}        % if too long for running head

\author{Matthias M\"oller \and
        Cornelis Vuik %etc.
}


\institute{M. M\"oller \at
              Delft University of Technology \\
              Delft Institute of Applied Mathematics\\
              Mekelweg 4, 2628 CD Delft, The Netherlands\\
              Tel.: +3115 27 89755\\
              Fax:  +3115 27 87209\\
              \email{m.moller@tudelft.nl}
           \and
           C. Vuik \at
              Delft University of Technology \\
              Delft Institute of Applied Mathematics\\
              Mekelweg 4, 2628 CD Delft, The Netherlands\\
              Tel.: +3115 27 85530\\
              Fax:  +3115 27 87209\\
              \email{c.vuik@tudelft.nl}
}

\date{Received: date / Accepted: date}

\maketitle

\begin{abstract}
Quantum computing technologies have become a hot topic in academia and industry receiving much attention and financial support from all sides. Building a quantum computer that can be used practically is in itself an outstanding challenge that has become the 'new race to the moon'. Next to researchers and vendors of future computing technologies, national authorities are showing strong interest in maturing this technology due to its known potential to break many of today's encryption techniques, which would have significant and potentially disruptive impact on our society. It is, however, quite likely that quantum computing has beneficial impact on many computational disciplines.

In this article we describe our vision of future developments in scientific computing that would be enabled by the advent of software-programmable quantum computers. We thereby assume that quantum computers will form part of a hybrid accelerated computing platform like GPUs and co-processor cards do today. In particular, we address the potential of quantum algorithms to bring major breakthroughs in applied mathematics and its applications. Finally, we give several examples that demonstrate the possible impact of quantum-accelerated scientific computing on society. 

\end{abstract}

\section{Introduction}
\label{sec:intro}

\emph{Quantum computing technologies} have become a hot topic that nowadays receives a lot of attention from researchers in academia as well as R$\&$D departments of the global players in computing. Intel \cite{Intel-Sep2015}, for instance, plans to invest about \$50 million over the next 10 years into research on quantum computing together with the Dutch research center QuTech\footnote{\url{http://qutech.nl}} that is affiliated with Delft University of Technology, while IBM\footnote{\url{http://www.research.ibm.com/quantum/}} builds on more than three decades research effort in this field and offers a cloud service to let students and researchers get practical 'Quantum Experience'.

It is clear that quantum computing has become the new 'race to the moon' pursued with national pride and tremendous investments. For instance, the European Commission \cite{EC-May2015} is planning to launch a \euro1 billion flagship initiative on quantum computing starting in 2018 with substantial funding for the next 20 years. This is already a follow-up investment in addition to the \euro 550 million that have already been spent on individual initiatives in order to put Europe at the forefront to what is considered the \emph{second quantum revolution}. While the first quantum revolution started in the early 1900s with the achievements of Plank, Bohr, and Einstein leading to a theoretical understanding of the behaviour of light and matter at extremely small scales, it is now considered timely to bring the technology to the next maturity level and build real quantum computers in order to exploit their theoretical superiority over today's classical Von-Neumann computers in practical applications.

\subsection{The past: Digital computer revolution}

Going back in history, the world's first \emph{programmable, electronic, digital computer}, the Colossus, was build by the research telephone engineer Tommy Flowers and used between 1943--1945 by British code breakers in Bletchley Park to decrypt and read secret messages of the German military during World War II. Another pioneer in this field, the Atanasoff-Berry computer, developed between 1937--1942 by John Vincent Antanasoff and Clifford Berry, should not go unnoticed. It deserves the credit of being the world's first electronic digital computer but is was not programmable and only designed to solve linear systems of equations. Next to Colossus, other computing machines like the U.S.-built ENIAC were designed during WWII to break decrypted messages. It took another 20 years before the first commercially available desktop personal computer, the Programma 101, was offered by Olivetti in 1964 at a regular price of \$3,200 which would correspond to \$20,000 today. The P101 made use of the techniques of its time, transistors, diodes, resistors and capacitors, and was used, e.g., by NASA to plan the Apollo 11 landing on the moon. It took another decade before the advent of microprocessors significantly reduced the costs of personal computers and made them a product for the masses. Further improvements in semiconductor and microprocessor technologies made it finally possible to significantly reduce the size and costs of integrated circuits and integrate all components of a computer into systems-on-a-chip bringing software-programmable computers for \$20 per device.

\subsection{The present: Quantum computer revolution}

Over the last decades, quantum technology has been an exciting toy for scientists but it still has to demonstrate its usefulness in practice. Frankly speaking, industrial interest and long-term investment in quantum hardware and software development can only be achieved if the overall benefits outweigh the immense costs of building and operating quantum computers and their infrastructure as well as developing quantum algorithms and, finally, applications for realistic problem sizes.

It is not a coincidence that the strongest interest in building practically usable quantum computers is largely motivated by their potential to break public-key cryptography schemes such as the widely used RSA scheme \cite{RSA1978}. The theoretical superiority of quantum computers in this particular discipline is based on Shor's quantum algorithm \cite{Shor1994} for the efficient factorization of large integer numbers into prime factors in polynomial time, whereas the most efficient classical algorithms require sub-exponential time. Appendix~\ref{appendix:complexity} gives a brief overview of the different complexity classes. Variants of the Rivest-Shamir-Adleman (RSA) encryption are used everywhere, for instance, for making secure connections to the Internet, sending text messages between mobile phones and email programmes and for signing contracts and official documents digitally. It is clear that the ability to read and possibly modify encrypted data and communication is most tempting for intelligence services and hackers alike, thus justifying research on quantum computers and algorithms for this purpose alone. It is, however, not completely unthinkable that quantum computers, like personal computers since the 1980s, will become available for the masses once the technologies for manufacturing and operating quantum hardware has matured and the total cost of ownership have reached an economically acceptable level. That said, we believe that the most probable scenario will be quantum computing as a service as it is already offered by IBM through its "Quantum Experience" servive \cite{IBM-Q}.

\subsection{The possible future: Quantum-accelerated computing as a service}

A common challenge of most of today's quantum devices is the need for extremely low operating temperatures near absolute zero, which suggests quantum computing as a cloud service as most promising business model to bring this technology to the end-users. However, this immediately raises the question about the reliability of results received from a quantum computer in the cloud when the communication takes place over an Internet connection that can be decrypted by other quantum computers.

Technology breakthroughs like the Transmon cryogenic 5-qubit devices \cite{Versluis2016} have heralded the era of practical quantum computers. Researchers worldwide are now focussing on maturing the mass production of multi-qubit devices so as to enable the construction of large-scale quantum computers with millions and billions of qubits \cite{Lekitsch2017}, which will be necessary to solve real-world problems. It is, however, equally important to create a quantum ecosystem \cite{Fu2016} consisting of a standardized quantum programming language \cite{Balensiefer2005}, compilers and debuggers \cite{JavadiAbhari2015}, and a quantum hardware abstraction layer \cite{Brandl2017} that allows to compile a single quantum program for different target quantum hardware platforms as it is common practice for classical computers. Furthermore, quantum computers need extra effort to detect and correct errors since all qubit technologies available today are very fragile and prone to errors.

In this article we describe possible scenarios of how the advent of practical large-scale quantum computers \emph{can} revolutionize scientific computing in the next decades. We thereby leave aspects of quantum hardware and the manual realization of quantum algorithms out of consideration and focus on quantum computers as software-programmable computing devices that enable the development, simulation, testing and analysis of device-independent quantum algorithms. It is our strong belief that quantum computers will not exist as stand-alone machines but need to find their niche in the global computing landscape. The future of scientific computing and quantum computing is, of course, not predictable. We therefore sketch a thinkable scenario that would maximise the impact of quantum computing on scientific computing, namely, quantum-accelerated computing brought to the end-user as a cloud service.

The remainder of this article is structured as follows: In Section~\ref{sec:sc} we briefly outline the current state of the art in scientific computing and continue with describing the challenges faced by future computing hardware in Section~\ref{sec:hardware}. Section~\ref{sec:principles} gives a very brief introduction into the principles of quantum computing to prepare the reader for the discussion of known quantum algorithms in Section~\ref{sec:algorithms}. The potential impact of quantum computing on computational sciences is sketched in Section~\ref{sec:impact} followed by a short outline of possible long-term quantum-enabled applications in Section~\ref{sec:applications}.

\section{Scientific Computing}
\label{sec:sc}

\emph{Scientific computing} is a rapidly growing multidisciplinary field that uses advanced computer simulation technologies to analyse complex problems arising in physics, biology, medicine, civil engineering, electrical engineering, aerospace engineering, social sciences and humanities to name just a few. Scientific Computing is nowadays also called the third pillar of scientific research, next to experimental and theoretical science. We observe that the range of applications becomes broader and broader. It started with Computational Fluid Dynamics, Computational Physics, Computational Chemistry,  and nowadays there is hardly a scientific field without a computational variant. Some examples are: Computational Finance, Computational Traffic Models, Computational Social Sciences and many more. One of the reasons is the enormous speedup in computer power and algorithmic performance. It is already possible to use advanced algorithms to simulate a fluid flow on a mobile phone.

Scientific computing is nowadays widely used in many disciplines, e.g., to
\begin{itemize}
\item predict and optimise the behaviour of new products such as diapers, vacuum cleaners, cars and aircrafts long before the first prototype is constructed;

\item predict, optimise and orchestrate the interplay of smart manufacturing devices such as, e.g., multi-robot systems as they are used in automotive industry;

\item predict and optimise the properties of novel materials such as complex composite materials or, only recently, graphene by controlling the creation process;

\item enable big data and predictive risk analysis in, e.g., flood, weather and epidemiological forecasting, emergency evacuation planing, and high-frequency trading;

\item provide deeper insight and theoretical understanding of complex problems such as the existence of black holes and the nature of dark matter \cite{NASA-June2015}, which are difficult or even impossible to study by experiment.
\end{itemize}

To judge the impact of Scientific Computing it is good to have a rough idea of how this is implemented for a real application. Let us consider the prediction of water levels in the North Sea (which is very important for the Netherlands). First a physical model of the water velocities and water height has to be made. The well known Navier-Stokes equations are a good start, but very hard to solve. So using a number of plausible assumptions a simplified model, the Shallow Water Equations, is formulated. Although these equations are easier to solve it is impossible to determine the solution in an analytical way. This means that a numerical model has to be made. Again a number of assumptions are used to derive a numerical model that has a solution which is computable and is a good approximation of the solution of the Shallow Water Equations. Finally, the numerical model has to be solved by a computer. Efficient algorithms, who have good approximation properties and are well suited to be implemented on modern hardware have to be used to compute the approximate solution. Then the results of all modelling and approximating activities have to be compared with water height measurements done in the North Sea.

Due to the tremendous increase in computer power (factor one million) and the huge increase in efficiency of the algorithms (also a factor one million) we are now able to simulate more and more complex phenomena. A societal danger is that the results of the approximation are judged as 'the true solution'. In our example we have seen that many assumptions and approximations are done so in problems where for a number of scenarios the approximations can be compared with measurements we can trust the  results, but for complicated and new applications the results should be interpreted with care. Are the assumptions valid? What is the effect of guessed coefficients? How large are the approximation and rounding errors? etc. It would be much better if not only a result is given but also a realistic error estimate is specified. In many simulations this is not done, so the quality of the results can not be judged. This is one of the dangers by developing more advanced mathematical models and more powerful computers that the results are interpreted as the truth, whereas for all Scientific Computing results the interpretation should be done in a critical way.

In what follows we briefly address important milestones in the historical development of scientific computing both from a hardware and software perspective and give some outlook on possible future technology trends in this field.

\subsection{Scientific computing from a hardware perspective}

\begin{figure}
\centering
\includegraphics[width=0.95\textwidth]{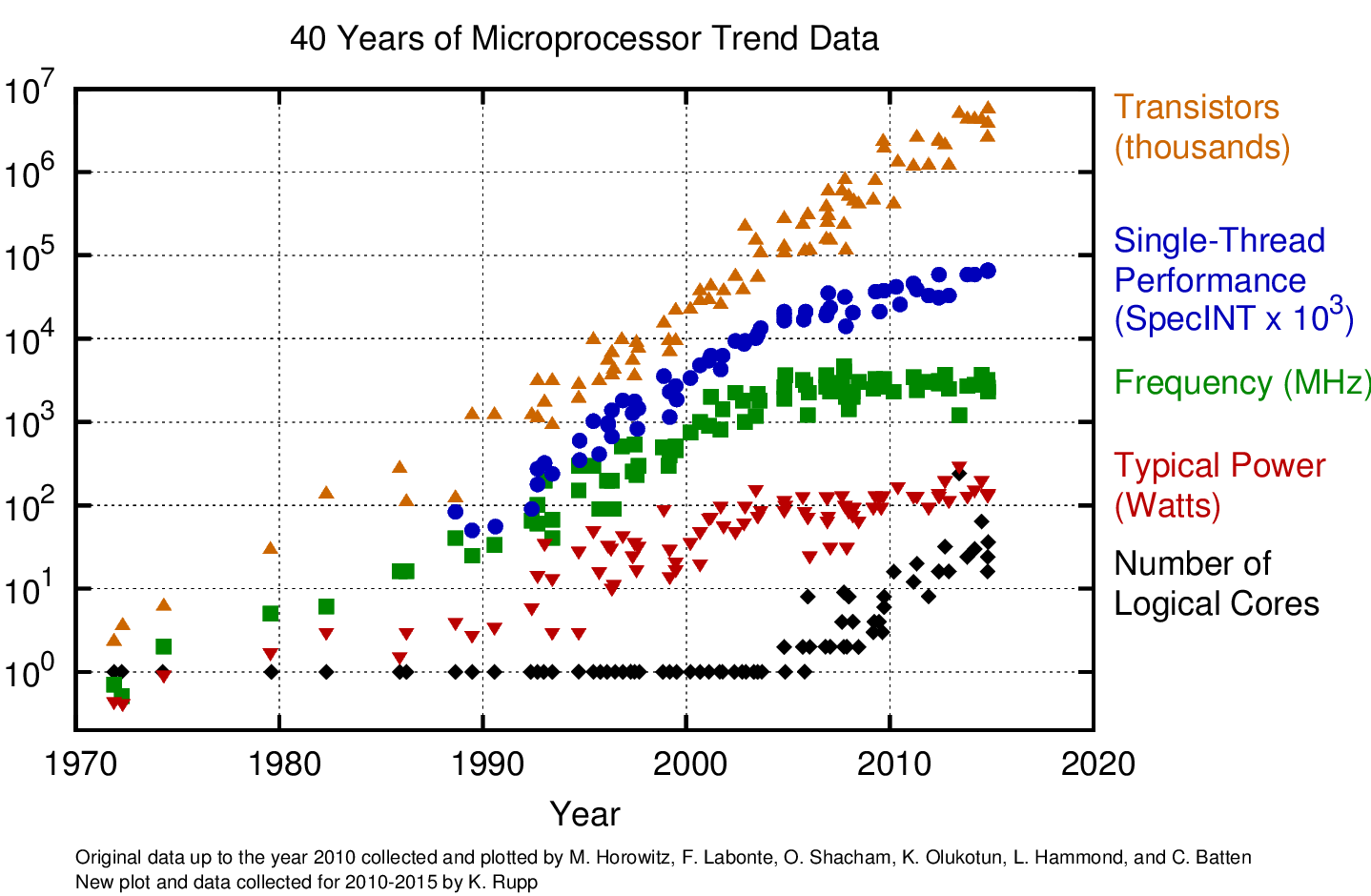}
\caption{40 years of microprocessor trend data \cite{Rupp2015}.}
\label{fig:microproc_trends}
\end{figure}

In the early days of scientific computing, parallel computers were very expensive and rarely available so that it was common practice for decades to develop sequential algorithms and implement computer programs for single-core machines. With each new hardware generation the performance of computer programs increased due to the increase of the CPU clock speed. This free-lunch strategy was strongly supported by major chip and computer system vendors until about 2005, when CPU clock speeds reached the 4 GHz barrier (Fig.~\ref{fig:microproc_trends}). Significantly increasing the clock speed beyond this barrier would require enormous effort for cooling the processor to prevent spurious malfunctioning and even permanent hardware damage from overheating.

\begin{figure}
\centering
\includegraphics[width=0.95\textwidth]{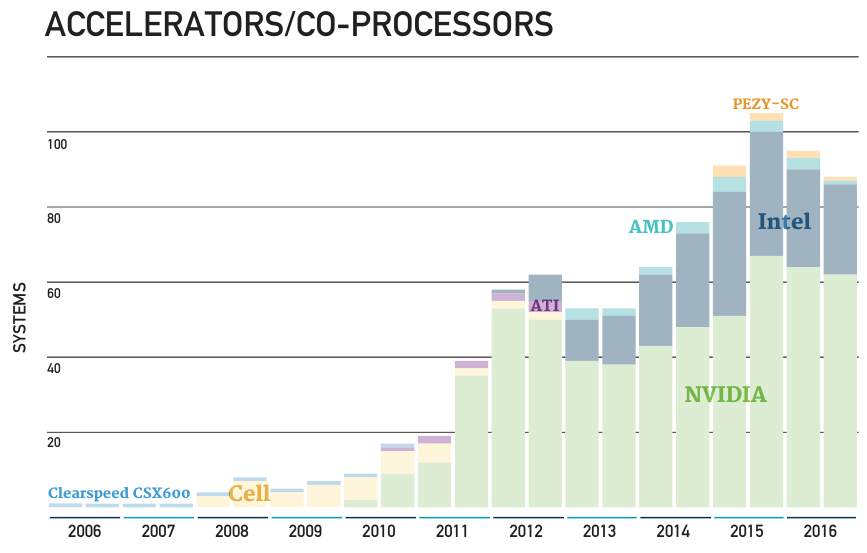}
\caption{Use of accelerators/co-processors in Top500 super-computers (TOP500 Nov 2016) \cite{Top500-Nov2016}.}
\label{fig:top500_accelerators}
\end{figure}

Since then, scientific computing has experienced a drastic paradigm shift from chasing ultimate single-core performance towards \emph{parallel high-performance computing} (HPC). Hardware vendors started to flood the market with cheaply available multi-core CPUs and many-core accelerator cards. So-called programmable general-purpose graphics processing units (GPGPUs) and dedicated co-processor devices like Intel's Xeon Phi have brought parallel computing to the masses thereby establishing the \emph{era of accelerated computing}. The key idea of accelerated computing is to offload those parts of the code that are  computationally most expensive and at the same time well-suited for parallelisation from the CPU,  termed the host, to the accelerator device. The host together with its accelerator device(s) forms the compute node. In this scenario, inherently sequential parts of the application and code that hardly benefits from parallelism are executed on the host, which moreover orchestrates the interplay of accelerator devices among each other and with the CPU and manages communication with other nodes. One fifth of the Top500 \cite{Top500-Nov2016} world's fastest supercomputers in 2015/2016 extracted their computing power from accelerator technologies (Fig.~\ref{fig:top500_accelerators}). 

However, the  offloading principle also has its downside. Since the raw compute power of chips is improving much faster than the speed of memory buses, component interconnects and network systems, the transport of data between the different memory layers and compute units as well as between hosts and accelerator devices has become the major bottleneck in data-intensive applications. The growing disparity of speed between compute and memory units is known as the \emph{memory wall} and it is nowadays one of the major bottlenecks in computer performance.

A new trend in scientific computing that aims at overcoming the memory-processor communication bottleneck is the rediscovery of reconfigurable hardware, e.g., Field Programmable Gate Arrays. FPGAs make it possible to design algorithms at hardware level thinking in terms of dataflow diagrams rather than control loops and function calls. The advent of general-purpose reconfigurable hardware once more requires a radical paradigm shift from traditional control-flow computing towards \emph{spatial computing} using for instance hybrid CPU-FPGA approaches like the Saturn 1 Hyperscale Server \cite{SMC-Sep2011} or Maxeler’s Multiscale Data-Flow Engines \cite{Pell2012}, which start to become accepted as reconfigurable HPC devices in the scientific computing community. 

An even more radical emerging technology is computing-in-memory \cite{Hamdioudi2015}, which aims at eliminating the central system bus as being the major performance bottleneck in today's computer systems completely. In short, the splitting between a central processing unit (CPU) and a hierarchy of memory tiers (Cache, RAM, storage) with relatively slow interconnects is abandoned in favor of a huge memristor-based memory pool with many small processing units located next to the storage cells on the die. Despite the early stage of this new technology, HP Enterprise has taken up the concept of memory-driven computing in their proof-of-concept realization of The Machine \cite{HPE-Nov2016}, which, in May 2016, has been expanded to a 160 terabyte single-memory computer.

In light of the above, one might come to the conclusion that the rank growth and diversity of ever new emerging technologies has never been as dynamic and widespread as today. However, exotic hardware architectures like, e.g., The Hypercube \cite{Millard1975}, which never became a commercial success story, existed at all times. In our opinion, the main difference today is the early availability of novel technologies to a broad community, which is largely made possible by cloud services.

The main findings from this historical review of hardware developments are:
\begin{itemize}
\item With current technology, further performance gains can only be achieved by the more effective exploitation of parallelism and by developing strategies to overcome the memory wall rather then by increasing single-core performance.
\item Future HPC systems are likely to become much more heterogeneous and massively-parallel systems with easier access for end-users enabled by cloud services
\end{itemize}

\subsection{Scientific computing from a software perspective}

With the advent of parallel computing as mainstream technology, software developers were forced to rewrite their codes basically from scratch making use of parallel computing technologies in order to benefit from improvements in hardware performance. However, the variety of parallel programming models (e.g., shared memory, message passing), parallelism strategies (e.g., instruction-level parallelism, task parallelism, data parallelism) and application programming interfaces (API) and languages makes choosing long-term investment-proof strategies that will extend to novel hardware platforms a challenging task. In many cases, the personnel costs for porting large scientific codes to new hardware architectures exceed the acquisition costs of the hardware by orders of magnitude, not to speak of the delay in scientific advancements.

The scientific computing community and HPC technology vendors have recognised the urgent need for developing novel types of meta-programming techniques to allow scientists to focus, again, on investigating their primary research questions and not wasting their time on repeatedly rewriting application codes for each new hardware generation. Next to the established parallel programming interfaces OpenMP \cite{OpenMP} and MPI \cite{MPI} new standards like OpenCL \cite{OpenCL}) have emerged with the ambition to provide device- and vendor-independent software frameworks for writing reusable code that runs on various types of heterogeneous platforms including CPUs, GPUs, and FPGAs. It is, however, still part of the responsibility of the application developer to design and implement the code in such a way that it respects the characteristics of the concrete hardware platform to achieve good performance, and thus, the vision of a fully device-independent abstract programming model remains wishful thinking.

An exception to this shift towards unifying frameworks is the CUDA toolkit \cite{CUDA}, which is a vendor-specific framework for enabling GPU-accelerated computing. Since the initial release of the CUDA software development kit in 2007, NVIDIA kept on enriching its capabilities by continuously adding highly optimized libraries that provide ready-to-use solutions for a broad range of scientific computing applications thereby attracting researchers from evermore disciplines. The lesson to learn from this very successful business model is that the acceptance of novel hardware architectures increases with the  availability of rich software stacks and the ease of access to hardware, e.g., through cloud services and  academic discount programs.

Another interesting trend is the advent of multi-platform accelerator libraries \cite{ArrayFire,Demidov2013}, which offer essential core functionality like fast linear algebra and solution routines under a unified API. It is the natural response to the fact that the largest group of researchers in the field of scientific computing are end-users of accelerator technologies and, thus, they are mainly interested in quickly developing solutions to their research questions rather then experimenting with the latest hardware developments. 

In line with this trend towards unifying device-independent application development frameworks is the appearance of middleware libraries, which allow application programmers to develop code in a device-independent kernel language that is compiled into compute kernels at run-time \cite{Medina2014} or to express common parallelisation patterns like forall-loops using device-independent meta-programming techniques \cite{Edwards2014}.

In our opinion the main drivers for the trends described above are the huge advancements in software technology like just-in-time compilation and meta-program\-ming techniques and the movement towards open-source software and open collaboration enabling synergy effects across the boarders of hardware and software vendors. Remarkably, most compiler vendors offer no-charge community editions of their premium products to assure their portion in the highly competitive HPC market.

The main findings from the review of recent software developments are:
\begin{itemize}
\item Device- and vendor-independent open standards and middleware software make parallel computing and accelerator technologies better accessible for end-users.
\item Community-based development of open-source software and the offering of professional software products free of charge has become a strong trend.
\item End-users are used to computer hardware being shipped with sophisticated software stacks and will expect this comfort from novel architectures as well.
\end{itemize}

\section{Challenges and strategies for future computing hardware}
\label{sec:hardware}

The fastest supercomputer in the Top500 list from November 2016 \cite{Top500-Nov2016} is the Sunway TaihuLight running at the National Supercomputing Center in Wuxi, China. It is equipped with 1.31 petabyte of main memory and has a maximum performance of 93 petaflops (a petaflow is $10^{15}$ floating-point operations per seconds) measured for the established Linpack benchmark, thereby exploiting $74\%$ if its theoretical peak performance of $125.4$ petaflops. This test consumed 15 megawatts of electrical power.

Despite these impressive figures, researchers worldwide make strong efforts to break the exascale barrier, that is, $10^{18}$ floating-point operations per second by the years 2020-2023. The main scientific and technological challenges that need to be addressed to make this dream come through are as follows \cite{ASCAC-2010,ASCAC-2014}:
\begin{itemize}
\item \emph{Reduction of power consumption.} Scaling today's computer technology to the exaflop level would consume more than a gigawatt of power for a single exascale system. To generate this amount of power requires about 400 wind turbines assuming an average capacity of 2.5 megawatt. A reduction in power requirement by a factor of at least 100 is thus needed to make exascale computing economical.

\item \emph{Coping with run-time errors.} Scaling today's technologies, exascale systems are expected to have approximately one billion processing elements. As a result, the frequency at which hardware errors occur will possibly increase by a factor of 1000 yielding Mean Time To Interrupts (MTTI) of 35-39 minutes \cite{Bergmann2008} for an exascale system. Thus, timely error detection and correction becomes more difficult.

\item \emph{Exploiting massive parallelism.} It is already a great challenge to effectively exploit the computing power of today's petaflop systems. In \cite{Dongarra2016}, Dongarra reports a sustained performance of 30-40 petaflops (24-32\% of the theoretical peak performance) for an explicit global surface wave simulation and only 1.5 petaflops (1.2\% of the theoretical peak performance) for a fully-implicit nonhydrostatic dynamic solver both running on about 8 million cores, that is, close to the full system scale. Thus, new concepts and programming paradigms are required to make better use of the immense raw compute power of future exascale systems.

\item \emph{Efficient data movement.} The movement of data between processors and memory as well as between processing nodes is the most critical barrier towards realizing exascale computing. Movement of data over long distances, e.g., through the complete system, requires a lot of energy not to speak of the time it takes to propagate information. Photonics offers a potential solution to reduce the energy consumption by a factor of 100 over electronic interconnect technology.
\end{itemize}

To address the above challenges in the coming years, CEA (Alternative Energies and Atomic Energy Commission) in France and RIKEN in Japan are committed to building energy-efficient ARM-based supercomputers \cite{Katsuya2017}, whereas the U.S. Department of Energy (DOE) plans to bring two exascale machines to fruition by 2023 most probably based on accelerators cards \cite{Feldman2016}. It needs, however, groundbreaking new approaches to pave the way for the future of scientific computing beyond exascale.

Quantum computing with its unique concept of quantum parallelism bears the potential to bring this revolution in scientific computing in the long run.

\section{Principles of quantum computing}
\label{sec:principles}

This section gives a brief overview of quantum principles helpful to recognize the possible impact of quantum computing on the future of scientific computing and the obstacles that need to be mastered. A more formal description is given in Appendix~\ref{appendix:qc}.

\subsection{Qubits and quantum circuits}
Bits, registers and logic gates are the basic building blocks of classical computers. Information is encoded as a sequence of bits, whereby established standards exist for storing, e.g., characters by the ASCII standard \cite{ASCII} or single- and double-precision floating-point numbers by the IEEE745 standard \cite{IEEE745}. For instance, the letter 'A' has ASCII code $\left.65\right|_{10}$ (in decimal representation), which is converted to the 8-bit sequence $\left.01000001\right|_2$. The advent of novel computer architectures has lead, however, to ever new ways of representing information aiming at narrowing the memory footprint of data by using half-precision intrinsics since CUDA 7.5 \cite{CUDA} or reducing the complexity of arithmetic logic units by fixed-point arithmetic on FPGAs. To prevent wild growth and incompatibility issues, committees worldwide strive to standardize new formats, e.g., half-precision floating-point numbers in the IEEE745-2008 standard, and compiler vendors make an effort to include them into their tools.

Such global standards do not yet exist for quantum computers so that the task of encoding input and output data is left to the quantum algorithm programmer. However, an efficient encoding of data is most crucial for efficient quantum algorithms since any algorithm that needs to read an input of length $n$ (and writes an output of the same length) cannot have overall time complexity better than linear in $n$ even if the actual 'processing' of the data once read into the quantum register takes time, say, $\mathcal{O}(\log n)$. As we are about to see in Section \ref{sec:algorithms:linsolve} this might even require the reformulation of the problem from writing out the raw solution (e.g., a vector of length $n$) to seeking a derive quantity, e.g., the sum of all vector entries (a scalar quantity).

Despite the lack of standardization, the concept of bits, registers and gates carries over to quantum computing with the exception that a quantum bit (termed \emph{qubit}) does not store the binary value $0$ or $1$ but holds a \emph{superposition} of all possible \emph{states} in-between. The conversion to one of the binary values (more precisely, the pure or basis states) is termed \emph{measurement} and it destroys the superposition of states.

The concept of superposition of states and the role of measuring is best illustrated by Schr\"odinger's famous thought experiment. A cat is placed in a steel box along with a Geiger counter, a vial of poison, a hammer, and a radioactive substance. The decay of the radioactive substance is a random process and, hence, it is impossible to predict when it will happen. Once it does happen, the Geiger counter will detect this and, according to Schr\"odinger's setup, it will trigger the hammer to release the poison, which will finally lead to the cat's death. However, it is not before an observer opens the steel box that he or she knows whether the cat is still alive or dead. Until this moment of measuring the cat is in a superposition state between life and death.

\subsection{Quantum parallelism and no-cloning principle}

The addition of two qubits yields a new state, also in superposition. The mathematical details of how to compute this state following simple linear algebra rules are given in Appendix~\ref{appendix:qc:qubits}. It is the superposition of states that makes quantum computing so powerful. Consider a set of qubits, a quantum register, where each qubit holds a superposition of states. That way, the quantum register in some sense holds all possible configurations of input data simultaneously. Let us perform Schr\"odinger's thought experiment with $n$ cats in $n$ separate boxes at the same time so that the measurement can yield $n$ dead or $n$  living or any combination of $k$ dead and $n-k$ living cat in-between. In other words, a single application of the quantum 'algorithm' to an $n$-qubit register calculates all possible $2^n$ combinations of states in parallel and it is the final measurement that converts the superposition of states into a definite answer. 

This feature of quantum computing, termed \emph{quantum parallelism}, is unparalleled in classical computing which can only process one combination of input data at a time and would require $2^n$ runs. However, $2^n$ individual classical computations yield the exact output to all possible combinations of input data from which the optimal value can be selected. In contrast, the outcome of the measuring procedure at the end of a single run of a quantum algorithm is a 'randomized' selection from the set of all possible solutions. Quantum algorithms therefore require special tricks that enhance the likelihood of measuring the desired solution and not just a random choice. It is this special type of quantum parallelism that \emph{can} lead to significant gains in efficiency provided that the quantum algorithm makes full use of it and has appropriate techniques to measure the desired output with high probability.

In addition to efficiency considerations it should be noted that classical divide-and-conquer strategies frequently used in scientific computing lack a quantum counterpart. For instance, the \emph{no-cloning principle} states that is it impossible to make a perfect copy of a qubit or quantum register without destroying the superposition state of the source. Further so-called no-go theorems have a huge influence on the way quantum algorithms must be designed. As an example, consider the simulation of water flow in the North Sea. A common practice in solving such huge problems, which exceed the memory capacities of a single computer, is to split the problem into many small sub-problems and distribute them to multiple computers, where they are solved in parallel. A core ingredient to domain decomposition techniques of this type is the ability to exchange information between different computers, that is, to copy data from one sub-problem to another. This is, however, impossible to achieve on quantum computers due to the no-cloning principle. In conclusion, many well-established classical concepts will require a complete redesign if they make use of concepts that violate one or more quantum no-go theorems.

\subsection{Reversible computing}
In most of today's computers computer programs are realized by logical gates like logical conjunction ($\land$) and disjunction ($\lor$), which map two  Boolean input values into a single Boolean output value. For the logical conjunction gate, the output value is true if and only if both input values are true ($1\land1=1$). On the other hand, it is impossible to derive the values of the two input values by just knowing that $a\land b=0$. In other words, the application of the logical conjunction is not reversible.

Quantum gates \emph{are}, however, reversible thanks to the unitary property of the transformation matrices. This means that any quantum circuit can be reversed by applying the sequence of  'inverse' quantum gates in reverse order to the output state vector.

Reversible computing has another interesting implications next to the possibility of 'undoing' algorithms. As shown by Landauer \cite{Landauer1961}, the erasure of a single bit of information requires a minimum amount of energy. Modern computer chips possess billions of irreversible logic gates leading to unwanted heat production. A modified chip design that is only based on reversible classical logic gates would reduce the energy consumption of computers. Since each input channel would be associated  with its unique output channel no bit would be erased, and hence, 'no' energy would be dissipated during computation. Only the initialization of registers and the storage of the computed answer would require some energy. 
As first noted by Landauer \cite{Landauer1961} and later refined by others \cite{Bennett1973,Fredkin1982,Lecerf1963}, any irreversible computation can be simulated by a reversible circuit. However, reversible computing is not yet practical.

\subsection{Application of reversible computing}

Postulating that quantum computers can bring reversible computing into practice, a couple of applications would immediately benefit \cite{Perumalla2013}. Debugging computer programs in forward and backward mode, that is, allowing the programmer to 'undo' steps is quite expensive in irreversible computing since intermediate results need to be stored. Reverse-mode debugging would be much simpler in reversible computers. A similar problem arises in reverse mode algorithmic differentiation (AD), which is a computational approach to quantify the sensitivity of the output of an algorithm to changes in its input values. AD is used in adjoint-based shape-optimization but the costs for storing all intermediate results are quite demanding.

The main findings from this short review of quantum principles are:
\begin{itemize}
\item Quantum computing still lacks a standardization for encoding input/output data.
\item Quantum algorithm development is based on linear algebra, stochastics and complexity theory and has little to do with programming as we know it today.
\item Quantum parallelism will be most effective if quantum algorithms are designed from scratch rather then simulating classical algorithms by quantum circuits.
\end{itemize}

\section{Algorithmic aspects of quantum computing}
\label{sec:algorithms}

We would like to begin this section by dispelling the myth that quantum computing will be the ultimate tool in solving the world's biggest problems. It should be clear to everybody that quantum computers will not be efficient per se but that a smart combination of quantum hardware and software, the optimal integration into classical computer platforms and the use of adequate quantum algorithms is required to deliver considerable speed-ups over classical technologies. In what follows we give several examples of quantum algorithms that might become  essential building blocks in scientific computing applications, once quantum hardware has reached a maturity level that will allow the computation of realistic problem sizes and accuracy tolerances of practical relevance. The focus is placed on numerical simulation and optimization, thereby keeping the level of technical details to a minimum to make this section accessible also for readers with less profound mathematical background knowledge. For an extensive list of quantum algorithms in other computational disciplines the interested reader is referred to \cite{SAND2015,Montanaro2016,Omer2009}. Readers interested in the impacts and applications of these algorithms can jump to section \ref{sec:impact}.

\subsection{Quantum-accelerated linear solvers}
\label{sec:algorithms:linsolve}

One of the most basic problems in scientific computing is the solution of systems of linear equations $Ax=b$ where $A$ is an invertible $N\times N$ matrix and $b$ a vector of size $N$. The most naive solution of this problem is Gaussian elimination without exploiting the system's sparsity pattern and it runs in time $\mathcal{O}(N^3)$. If $A$ is $d$-sparse, that is, each row contains at most $d\ll N$ entries, then the runtime of classical algorithms still scales at least linearly in $N$. This also applies to any quantum algorithm if the desired output is the solution vector $x$ which requires time $\mathcal{O}(N)$ just for being written out. 

However, if the quantity of interest is a scalar value $x^\top M x$ for sparse matrix $M$ then quantum algorithms exist with run-time polynomial in $\log N$, $d$ and $\kappa$ given that matrix $A$ has a small condition number $\kappa=\|A\|\|A^{-1}\|$. The first quantum algorithm for solving linear systems of equations with sparse matrices has been developed by Harrow, Hassidim, and Lloyd, and therefore, it is referred to as the HHL algorithm \cite{Harrow2009} in literature. Since then, improved variants with better run-time have been proposed by Ambainis \cite{Ambainis2010} and, more recently, by Childs et al. \cite{Childs2015}. Estimating the value of $x^\top M x$ by classical algorithms requires still linear time $\mathcal{O}(N\sqrt{\kappa})$ so that, at least for small $d$ and $\kappa$, quantum algorithms provide exponential improvement.

In fact, asking for a scalar output instead of the complete solution vector is quite common in scientific computing. Many physical problems like the flow of water in the North Sea are modelled by systems of partial differential equations (PDEs), which need to be discretized both in time and space to turn them into sparse systems of (non-)linear equations with millions or even billions of unknowns. Engineers are typically not interested into the complete flow pattern but rely on scalar quantities of interest like the tidal range at a critical location to design, say, flood protection systems.

\subsection{Quantum-accelerated design optimization}
\label{sec:algorithms:opt}

Derived quantities of interest become even more important when it comes to computer-aided design optimization. A common task in the automotive, aerospace, and naval industries is to optimize the shape of cars, aircrafts, and ships with the aim to reduce, say, the drag coefficient, while at the same time improving the lift coefficient with the direction of improvement depending on the particular applications. Multi-disciplinary design optimization problems involve multiple of these target quantities. However, the main challenge comes from the many design parameters that need to be varied in order to optimize the shape, which can easily reach hundreds or thousands of degrees of freedom. The main computational costs often arise from the evaluation of the cost functional, that is, the numerical simulation run for a particular set of design parameters. Thus, a good metric of the overall computational costs is the number of queries to the cost functional triggered by the optimization algorithm.

Close to a minimal solution, the objective function can be approximated by a Taylor series thus leading to the problem of minimizing a positive-definite quadratic form. Classical algorithms cannot do better than $\mathcal{O}(d^2)$ queries \cite{Yao1975}, where $d$ represents the number of design variables. In contrast, it is possible to find the minimum of a quadratic form in only $\mathcal{O}(d)$ quantum queries \cite{Jordan2008} thereby exploiting the concept of superposition of states and resorting to an efficient quantum algorithm for estimating gradients \cite{Jordan2005}. It is even possible to cure the common problem of gradient-based optimization algorithms, namely to get trapped into local minima rather than finding the global minimal solution, by resorting to quantum  annealing \cite{Somma2007,Szegedy2004}.

\subsection{Quantum-accelerated integration and summation}
\label{sec:algorithms:int}

For the numerical solution of partial differential equations (PDEs) the differential operators are typically approximated by discretization techniques like the Finite Difference, Finite Volume, or Finite Element Method (FEM), thereby involving summation of data and/or numerical integration. For the latter, quantum algorithms are known that provide quadratic speedup over classical ones \cite{Heinrich2002,Novak2001} showing their full potential for high-dimensional integrals \cite{Heinrich2003} as they occur for instance in Computational Finance. Unfortunately, research activities in this field have lost impetus, which might change once practical quantum computers become available making quantum summation and integration a building block for other quantum algorithms.

\subsection{Applications of quantum-accelerated linear solvers}

The HLL quantum algorithm \cite{Harrow2009} for solving linear systems of equations has been applied to various applications.  Clader et al. \cite{Clader2013} developed a preconditioned Finite Element Method (FEM) for solving electromagnetic scattering problems modelled by PDEs with polynomial speedup, whose theoretical analysis was later improved in \cite{Montanaro2016a}. For this application the subtle difference to the original HLL algorithm, where matrix $A$ is considered to be given as a function of the row number $r$ and the index $1\le i\le d$, is that in FEM matrix $A$ is constructed algorithmically by numerical integration.

Further applications of the HLL algorithm are related to the solution of large sparse systems of linear \cite{Berry2014,Berry2017} and nonlinear \cite{Leyton2008} differential equations, which play an important role in computational biology, e.g., predator-prey models, tumor growth and anti-angiogenic or radiation treatment, in computational neuroscience, e.g., models of the nervous system, and in other computational disciplines that focus on large but sparsely connected networks like, e.g., gas or power grids.  

\subsection{Challenges and potential of quantum algorithms}
\label{sec:algorithms:new}

The main difference between the quantum algorithms outlined above and, say, Shor's algorithm \cite{Shor1994} for factorizing a natural number $n\in\mathbb{N}$ into its prime factors is the size of the input data. Since $\log n$ qubits suffice to encode the input for Shor's algorithm, a quantum computer with $\sim 50$ qubits \cite{IBM-Qsystem} might already be of practical use. 

In contrast, computing meaningful results to the aforementioned applications requires possibly thousands or millions of qubits, thereby taking into account that up to 90\% of the quantum resources might be necessary to implement quantum error correction techniques \cite{Isailovic,Steane2007,Thaker2006}. This is, however, also a chance to strengthen interdisciplinary research. With the severe limitations of quantum hardware resources that can be expected to persist at least in the coming years it might be worthwhile to store data most efficiently, e.g., by using data compression techniques from coding theory. It might also be worthwhile to carefully analyze the number of qubits that is really needed to produce solutions with accuracies of engineering relevance. This might, in turn, stimulate a paradigm shift in classical computing from using IEEE754 floating-point arithmetic unconditionally towards adopting storage techniques with smaller memory footprint. Remarkably, this is a recent trend in accelerated computing, where the limited resource is the memory bandwidth rather than the size.

Reliable and efficient error correction is indeed one of the greatest challenges in quantum computing. Most classical techniques like repetition codes, that is, the multiple repetition of the data is ruled out by the no-cloning principle. Thus, specialized quantum error correction techniques \cite{Terhal2015} are required such as surface codes \cite{Hill2015}. Many classical fault-tolerance techniques rely on invariance checking, that is, the comparison of intermediate values with known reference data. For instance in algorithms like the Conjugate Gradient method, which is based on the idea of orthogonalizing a sequence of vectors step by step, this invariant can be checked explicitly for pairs of vectors. However, invariant checking is much more difficult to realise on a quantum computer since the direct measurement of intermediate states is impossible without destroying the superposition state thus preventing further computations.

As stated above, the mean time to interrupts might drop to minutes as in exascale computing thus making error correction and fault-tolerance an integral part of future computer codes. Classical computers are considered to be deterministic devices in most cases and the outcome of a deterministic algorithm is expected to remain unchanged over multiple runs. However, parallelization strategies like divide-and-conquer and asynchronous parallel computing break the rules of traditional mathematics. For instance, the sum of three numbers $a$, $b$, and $c$ might slightly vary due to round-off and cancellation errors depending on the order of accumulation, i.e. 
$$
\text{fl}(\text{fl}(\text{fl}(a)+\text{fl}(b))+\text{fl}(c))
\ne
\text{fl}(\text{fl}(a)+\text{fl}(\text{fl}(b)+\text{fl}(c))).
$$
In this sense, both classical and quantum computing fail to compute \emph{the} approximate solution even for a uniquely solvable problem but generate only one possible answer. This observation might trigger a paradigm shift towards uncertainty quantification of simulation results by default. In the ideal case, the ever increase of computer power will not be used to compute more complex problems and/or larger problem sizes with less and less reliability but to simulate the same problems but with a quantified error range, which might be of particular interest for engineers

Last but not least, the advent of practical large-scale quantum computers might change the way in which quantum algorithms are designed and analyzed. In most publications, the efficiency of quantum algorithms is assessed by a theoretical formal complexity analysis. In the analysis of the HLL algorithm it is crucial that the solution vector is not written out, which would lead to linear complexity. However, the overall time-to-solution of a quantum computer might still be much smaller (or larger) compared to a classical computer. At the end of the day, theoretically sub-optimal quantum algorithms might become presentable, if they have practical benefits.

The main findings of this section are as follows:
\begin{itemize}
\item The zoo of existing quantum algorithms offers potential for speeding-up the solution of many challenging scientific computing problems once practical large-scale quantum computers become available and technical obstacles are mastered.
\item Classical and quantum computing face the same challenges -- reliability and uncertainty of results -- which might be addressed in joint effort.
\end{itemize}

\section{Impact of quantum computing on scientific computing}
\label{sec:impact}

Note that the construction and maintenance of a quantum computer is difficult, very expensive, needs special buildings and expert knowledge. A danger is that only a limited number of institutes in the world have access to these powerful machines. This will hamper the development of modern solution tools and can give these countries, institutes, and universities a decisive lead in scientific computing. To mitigate this danger it is possible to make quantum computing available via cloud services. 

Another danger is that the unparalleled possibilities of future computers might lead to a blind trust in simulation results. Already with today's technology, scientific computing combined with mathematical modelling is a very strong tool to analyze many phenomena and predict effects of certain changes. Examples are the analysis of the spread of diseases or the prediction of temperatures due to climate change. Scientific computing becomes in this way an important source to society for understanding of and for policy decisions on such topics. However, all these models are only valid if the assumptions used in their derivation are satisfied. Furthermore, the predictions computed with the aid of these models are only meaningful if the input data are reliable. The increase in computing power will drive the development of more and more complicated, misleadingly termed detailed, models, which require more and more complex input data. To say it frankly, the accuracy of a mathematical model will not increase by replacing a single unknown parameter by a dozen of unknown parameters but it requires reliable (measurement) data to make the enhanced model meaningful. In any way, this trend towards more and more complex simulations will strengthen the 'trust' in scientific computing. A danger is that the computations \emph{are} correct but that the assumptions are not satisfied and/or the input data is not reliable, which makes that the prediction can only be used in a careful way. Another aspect is the fact that in quantum computing errors will always occur due to the quantum effect in the computations, which makes the interpretation of the results even more difficult. Therefore, for having trustful results new ways should be developed for the validation of results. In our opinion using quantum computing it should be required not only to give a final result but to also provide a robust error estimate.

Radically different programming models require significant changes in teaching. Nowadays, programming is considered a technical skill, that is considered simple enough to be taught only superficially but to broad masses of students. In fact, a predominant opinion at universities is that educating programming skills is just a requirement for demonstrating the applicability of numerical methods but not a discipline in itself. This is a pity that needs to be corrected since the full power of already today's supercomputers is only exploitable by a negligibly small part of students and of, unfortunately, even researchers. In order to establish quantum computing as mainstream technology the 'art of programming' must receive more attention, again, which might in turn strengthen the interest of researchers in classical HPC and the willingness to invest effort in developing hardware-aware algorithms.

Finally due to better and faster computations it may seem attractive to replace experiments by models and simulations. Although there can be a shift into more simulations and less experiments, it will always be necessary to validate the results of a quantum computing algorithms with carefully designed experiments.

\section{Societal applications for quantum computing}
\label{sec:applications}

In this section we summarize five applications of quantum computing. Some of them are already simulated by first-generation quantum devices as the D-Wave systems\footnote{\url{https://www.dwavesys.com/}}, whereas others are only foreseen to be simulated by emerging quantum computers.

\subsection{Green aircraft}

Big aircraft companies are working in developing and using quantum algorithms to predict the flow of air over a wing \cite{Tovey2017}. Using classical computers such simulations can take more than seven years of computing time. Quantum computers should be able to simulate all the atoms of air flowing over the wing using various angles and speeds in several weeks. This can enhance the modelling and optimization methods considerably, enabling the aircraft designers to develop robust and efficient aircraft with low noise and $CO_2$ emission in a much shorter period of time.

\subsection{Optimization in space applications}

In the NASA department Quantum Artificial Intelligence Laboratory (QuAIL)\footnote{\url{https://ti.arc.nasa.gov/tech/dash/physics/quail/}} research is done to assess the suitability of quantum computers for optimization problems that are of practical relevance for aerospace applications.

A start has been made by using the D-Wave Two\texttrademark quantum computer with a quantum annealing optimization method to optimize various applications ranging from optimal structures to optimal packing of payload in a space craft. One aspect, which is important to investigate, is the effect of numerical noise inherent to quantum computing which influences the final result. Other applications, which are considered in this laboratory, are quantum artificial intelligence algorithms, problem decomposition and hardware embedding techniques, and quantum-classical hybrid algorithms.

\subsection{Secure communication technology}

A well known application is quantum encryption. Currently used encryption methods can be easily broken by future quantum computers. The reason is that the security of the used encryption protocols is based on the fact that in order to break them a very time-consuming problem should be solved. Since the public keys are changed every week, this time is too short to break the code. Many new quantum algorithms are designed to provide secure communications after quantum computers become available that can break the current codes. A secure solution of the key exchange problem is quantum key distribution. Recently DLR \cite{DLR2013} has done a number of successful experiments to transmit a quantum key from a fast-moving object. The quantum data was sent from an aircraft to a ground station via a laser beam. These experiments show that encryption technology can be used with fast-moving objects. Furthermore, existing optical communications systems are able to transmit this information. 

\subsection{Flood predictions}

Many practical applications are based on flow of air, water or other liquids. The underlying model are the Navier-Stokes equations. Solving this type of equations in an efficient way is one of the most challenging problems in computational physics. Modelling turbulence for instance is one of the millennium problems that is not solved yet. In \cite{Mezzacapo2015}, a quantum simulator is developed, which is suitable for encoding fluid dynamics transport phenomena within a lattice kinetic formalism. The basis of this simulator comes from the analogies between Dirac and lattice Boltzmann equations. In \cite{Mezzacapo2015} it is shown how the streaming and collision processes of lattice Boltzmann dynamics can be implemented with controlled quantum operations. The proposed simulator is amenable to realization in controlled quantum platforms, such as ion-trap quantum computers or circuit quantum electrodynamics processors. This opens a large area of applications running from high-tension blood flow in the hearth, flow in industrial furnaces to the protection of low-lying countries for sea-water flooding.

\subsection{Medicine}

Quantum computing seems to be also suitable to model molecular interactions at an atomic level \cite{Diamandis2016}. Gaining insight into this process is of primary importance to develop new medicines or to understand various diseases. The future is that all 20,000+ proteins in the human genome can be modelled and the interaction with existing or newly developed drugs can be investigated. Again, this can help to lower the time to bring newly designed drugs to the patient. Using quantum computer simulations can be the way we design and choose our next generations of drugs and cancer cures.

\section{Conclusion}

In this article we shed some light on the possible impact of large-scale practical quantum computers on future developments in the field of scientific computing. First and foremost, quantum computers, quantum algorithms, and quantum principles are very different from all what we are used to know from classical computing based on digital circuits. However, classical computing also needs drastic changes to overcome its omnipresent limitations, namely, the memory wall, the energy wall, and the instruction-level parallelism wall. Knowledge transfer between both worlds might therefore be worthwhile. The quantum community can benefit from the long-term experience in classical computing with bringing novel architectures to the end-users. Manufacturers of conventional computers chips might, in turn, profit from quantum principles like reversible computing to improve their chip technology further.

In our opinion, quantum-enhanced scientific computing is an exciting new field that has the highest chances to become a game-changing technology if quantum hardware gets integrated into conventional HPC systems and used as special-purpose accelerators for those tasks for which efficient quantum algorithms exist. Approaches like quantum-accelerated cloud services are required to bring practical quantum computers to the stakeholders from industry and academia, which will help quantum computing as possible next-generation compute technology to pass the valley of death.

\begin{acknowledgements}
The authors would like to thank Koen Bertels and Carmen Almudever for very fruitful discussions on quantum computing technology. Moreover, we acknowledge the valuable feedback from the reviewer Pieter Vermaas.
\end{acknowledgements}

\begin{appendix}
\section{Complexity analysis of algorithms}
\label{appendix:complexity}
 
In theoretical complexity analysis one is mainly interested in the asymptotic complexity of an algorithm, which makes it possible to compare the complexity of different algorithms for solving the same problem. As an example, consider the task of reading an integer vector of length $n$ into computer memory. It is clear that each of the $n$ positions has to be visited at least once, and therefore, any read-in algorithm must have \emph{linear} complexity in the vector length. In a concrete implementation it might be possible to read-in two consecutive entries at a time, so that only $n/2$ elemental reads are required and the absolute wall-clock time halves. Nonetheless, the complexity of the algorithm is still \emph{linear} in the vector length.

\subsection{Bachmann-Landau notation}
The Bachmann-Landau notation, also termed the big $\mathcal{O}$-notation, has been introduced to simplify the study of asymptotic behavior of functions. Simply speaking, $f(n)=\mathcal{O}(g(n))$ for $n\to \infty$ means that the two functions $f(n)$ and $g(n)$ grow (or decay) equally fast in the limit. For example, $f_1(n)=3n^2$ and $g_1(n)=2n^2$ both grow quadratically ($f_1(n)=\mathcal{O}(g_1(n))$), whereas $f_2(n)=3n^3$ grows much faster than $g_2(n)=2n^2$, and hence, $f_2(n)\ne\mathcal{O}(g_2(n))$. The formal definition of the big $\mathcal{O}$-symbol reads as follows:
\begin{definition}
\label{def:landau}
Let $f$ and $g$ be two given functions. Then $f(n)=\mathcal{O}(g(n))$ for $n\to \infty$, if and only if there exist a positive constant $M$ and a number $n_0$ such that
$$
|f(n)|\le M |g(n)|\quad\text{for all}\quad n\ge n_0.
$$
\end{definition}

\subsection{Polynomial time complexity}
Let us consider the complexity of the Gaussian elimination algorithm (cf. Section~\ref{sec:algorithms:linsolve}) for solving linear systems of equations of the form $Ax=b$, where $A$ is an invertible $n\times n$ matrix and $x$ and $b$ are column vectors of size $n$. The asymptotic complexity of this algorithm is $\mathcal{O}(n^3)$, which implies that each of the $n\times n$ matrix entries is touched about $n$ times. A detailed analysis \cite{Farebrother1988} of the computational steps reveals that approximately $\frac23n^3$ arithmetic operations are required in a practical implementation.

Assuming that all arithmetic operations require a comparable amount of computing time (to be stashed by the big $\mathcal{O}$-notation), Gaussian elimination produces the solution vector $x=A^{-1}b$ in cubic polynomial time. More generally speaking, algorithms which solve the given problem in time $\mathcal{O}(n^k)$ for some constant $k$ are classified as \emph{polynomial time} algorithms. An alternative formalization reads $\text{poly}(n)=2^{\mathcal{O}(\log n)}$.

\subsection{Exponential time complexity}
Algorithms, which require time $2^{\text{poly}(n)}$ are classified as \emph{exponential time} algorithms implying that the time complexity grows exponentially with the problem size. For instance, the brute-force approach to solving a Sudoku puzzle by trying all possible combinations leads to exponential time complexity. Such extensive search of the solution space is a common strategy to solve combinatorial problems termed backtracking. In essence, for each empty position we guess an admissible number and proceed to the next empty position, thereby sequentially filling the puzzle in a particular order. Whenever we reach a dead end we backtrack to an earlier guess trying something else until we find a solution or conclude that the problem is not solvable once all possibilities have been explored unsuccessfully. Backtracking is a depth-first search strategy, which might end up trying all $6.67\times 10^{21}$ possibilities of admissible grids in the worst case.

\subsection{Sub-exponential time complexity}
Between the two aforementioned complexity classes lies the class of \emph{sub-exponential time} algorithms, which are formally characterized by time complexity equal to $2^{\mathcal{O}(n)}$. An alternative characterization of this class, which admits a more constructive interpretation reads as follows: If an algorithm solves the problem of size $n$ in time  $\mathcal{O}(2^{n^\epsilon})$ for all(!) $\epsilon>0$ then it has sub-exponential complexity. Going back to Definition~\ref{def:landau} this means that for all possible values $\epsilon>0$, we need to be able to find a (probably $\epsilon$-dependent) pair $(M_\epsilon,n_{0,\epsilon})$ of positive constants such that the time $T(n)\le M_\epsilon 2^{n^\epsilon}$ for all $n\ge n_{0,\epsilon}$.

\subsection{A final word on algorithmic complexity in practice}
It is our strong belief that the constants hidden behind the big $\mathcal{O}$-notation \emph{are}  relevant for practical applications. Given that building a universal quantum computer with $\sim 50$ qubits in the next few years \cite{IBM-Qsystem} is considered a major milestone, a practical complexity analysis for problem sizes approaching 50 might be more helpful for the coming decades. As thought experiment, consider the Gaussian elimination algorithm for the solution of a $6\times 6$ binary linear system $Ax=b$ with matrix $A\in\{0,1\}^{6\times 6}$ and vectors $x,b\in\{0,1\}^6$. In a na\"ive implementation this problem requires $6^2+2\cdot6=48$ bits for storing input and output data, which corresponds  to approximately $\frac236^3=144$ arithmetic operations. The solution of linear systems of equations can also be accomplished by combining  Strassen's algorithm ($\mathcal{O}(n^{2.807355})$) \cite{Strassen1969} or an optimized variant of the Coppersmith-Winograd algorithm ($\mathcal{O}(n^{2.3728639})$) \cite{LeGall2014} for matrix-matrix multiplication with a divide-and-conquer strategy based on block-wise inversion. However, to compete with the theoretically slower Gaussian elimination algorithm the constants 'hidden' in the big $\mathcal{O}$-notation must not exceed 0.9 and 2, respectively, rendering both approaches impractical for problem sizes of $n=6$.

\section{Principles of quantum computing}
\label{appendix:qc}

In what follows we give a brief description of quantum principles and their impact on scientific computing. For a thorough introduction into structured quantum programming the reader is referred to \cite{Omer2009}.

\subsection{Qubits and quantum circuits}
\label{appendix:qc:qubits}

Classical digital computers adopt a binary representation of information by a sequence of bits $b\in\{0,1\}$. The smallest possible unit in quantum computing is the so-called \emph{quantum bit} also termed \emph{qubit}. In contrast to classical bits, which can attain exactly one of the two possible states zero and one at a time, qubits are in a \emph{superposition} of both states
$$
\ket{\phi}=\alpha\ket{0}+\beta\ket{1},\quad \alpha,\beta\in\mathbb{C},\quad |\alpha|^2+|\beta|^2=1,
$$
where $\alpha$ and $\beta$ are probability amplitudes and $\{\ket{0},\ket{1}\}$ denotes the standard basis. When the state of qubit $\ket{\phi}$ is \emph{measured} in the standard basis, the probability of outcome $\ket{0}$ is $|\alpha|^2$ while the probability of outcome $\ket{1}$ is $|\beta|^2$. Thus, measuring of a qubit amounts to destroying the superposition of states and converting it into a classical bit that can only attain one of the two states zero or one at a time.

Quantum algorithms are realized by unitary transformations of \emph{state vectors}
$$
\ket{\phi}=\alpha\ket{0}+\beta\ket{1}\quad\to\quad 
\begin{bmatrix}
\alpha\\
\beta
\end{bmatrix}.
$$
Let $A$ be a unitary $2\times 2$ matrix, that is $(A^*)^\top=A^{-1}$, then the transformed qubit reads
$$
\ket{\phi^\prime}=\alpha^\prime\ket{0}+\beta^\prime\ket{1},\quad\text{where}\quad
\begin{bmatrix}
\alpha^\prime\\
\beta^\prime
\end{bmatrix}=
A
\begin{bmatrix}
\alpha\\
\beta
\end{bmatrix}.
$$
In the quantum circuit model of computing, the unitary matrices are associated with quantum gates, which form the basic building blocks for constructing complex quantum circuits just like classical logic gates do for conventional digital circuits. 

Consider as example the \emph{Hadamard gate} $$H=\frac{1}{\sqrt{2}}\begin{pmatrix}1&1\\1&-1\end{pmatrix}$$ which maps the two standard basis states into superposition states
$$
H\ket{0}=\frac{1}{\sqrt{2}}(\ket{0}+\ket{1}),\quad
H\ket{1}=\frac{1}{\sqrt{2}}(\ket{0}-\ket{1}).
$$
Upon measurement in the $\{\ket{0},\ket{1}\}$ basis, both states have equal probability to become either $\ket{0}$ or $\ket{1}$. Hadamard gates are frequently used for qubit initialization.

\subsection{Quantum parallelism}

A collection of multiple qubits is termed a \emph{quantum register}. In contrast to a classical $n$-bit register, which can only store a single value of the $2^n$ possibilities, an $n$-qubit quantum register holds a superposition of all $2^n$ possible classical states
$$
R=\ket{\phi_n}\ket{\phi_{n-1}}\dots\ket{\phi_0}.
$$
Quantum gates that act on $n$-qubit registers are described by unitary $2^n\times 2^n$ matrices. Due to the superposition of basis states, all possible $2^n$ input values are processed simultaneously within a single application of a quantum gate and, consequently, quantum circuit. In contrast, a classical digital circuit can only process single input value at a time and must therefore be run $2^n$ times. This unique property of quantum circuits is termed \emph{quantum parallelism} by the physicist David Deutsch and it forms the basis for the exponential performance boost expected from quantum computers.

However, quantum parallelism has two major antagonists: Firstly, a single run of the probabilistic quantum algorithm is insufficient since its outcome is mostly random. Thus, the quantum algorithm has to be run multiple times before, e.g., a majority vote can deliver the final result. This brings us to the second challenge. It is of course desirable to obtain the final result in much less than $\mathcal{O}(2^n)$ runs. Thus, the algorithm must incorporate mechanisms to amplify the probability of measuring the 'right' outcome, that is, the one that is closest to the desired solution of the problem.

\subsection{Entanglement and quantum no-go theorems}

Last but not least, quantum mechanics knows a special feature known as \emph{entanglement}. As an illustration, consider the 2-qubit register in the so-called Bell state
$$
\frac{1}{\sqrt{2}}(\ket{00}+\ket{11}),
$$
where the probability of measuring either $\ket{00}$ or $\ket{11}$ is $1/2$. If the two entangled qubits are separated and given to two independent observers at different locations, then if suffices to measure exactly one qubit to know the state of the other. This unique feature is considered one of the main reasons for quantum algorithms being more efficient than classical ones and it is used for instance in quantum teleportation. 

Entanglement of qubits must not be confused with copying states between qubits. In fact, the \emph{no-cloning principle} states that it is impossible to create an identical copy of an arbitrary unknown quantum state. Measurement is not an option since it would destroy the superposition state of the original qubit. Quantum information theory has more no-go theorems of this type, e.g, the no-communication and no-deleting theorem, which complicate the adaptation of classical algorithms to quantum computers.
\end{appendix}



\end{document}